\documentclass[reprint,amsmath,amssymb,aps]{revtex4-1}
\usepackage{mathptmx}
\usepackage[latin9]{inputenc}
\usepackage{verbatim}
\usepackage{units}
\usepackage{amsmath}
\usepackage{amsthm}
\usepackage{amssymb}
\usepackage{graphicx}

\makeatletter
\theoremstyle{plain}
\newtheorem*{fact*}{\protect\factname}

\usepackage{color}
\usepackage{dcolumn}
\usepackage{bm}
\usepackage{units}
\usepackage{xcolor}

\@ifundefined{showcaptionsetup}{}{%
 \PassOptionsToPackage{caption=false}{subfig}}
\usepackage{subfig}
\makeatother

\providecommand{\factname}{Fact}

\begin{document}

\preprint{APS/123-QED}

\title{Dynamical and Coupling Structure of Pulse-Coupled Networks in Maximum
Entropy Analysis}

\author{Zhi-Qin John Xu$^{1}$}

\thanks{zhiqinxu@nyu.edu}

\author{Douglas Zhou$^{2}$}

\thanks{zdz@sjtu.edu.cn}

\author{David Cai$^{1,2,3}$}

\affiliation{$^{1}$NYUAD Institute, New York University Abu Dhabi, Abu Dhabi,
United Arab Emirates,~\\
 $^{2}$School of Mathematical Sciences, MOE-LSC and Institute of
Natural Sciences,Shanghai Jiao Tong University, Shanghai, P.R. China,~\\
 $^{3}$Courant Institute of Mathematical Sciences and Center for
Neural Science, New York University, New York, New York, USA.~\\
 }

\date{\today}
\begin{abstract}
Maximum entropy principle (MEP) analysis with few non-zero effective
interactions successfully characterizes the distribution of dynamical
states of pulse-coupled networks in many experiments, e.g., in neuroscience.
To better understand the underlying mechanism, we found a relation
between the dynamical structure, i.e., effective interactions in MEP
analysis, and the coupling structure of pulse-coupled network to understand
how a sparse coupling structure could lead to a sparse coding by effective
interactions. This relation quantitatively displays how the dynamical
structure is closely related to the coupling structure. 
\begin{description}
\item [{PACS numbers}] 89.70.Cf, 87.19.lo, 87.19.ls, 87.19.ll 
\end{description}
\end{abstract}

\pacs{Valid PACS appear here}

\maketitle
Binary-state networks---each node in one sampling time bin is binary-state---arise
from many research fields, e.g., gene regulatory modeling and neural
dynamics \cite{mirollo1990synchronization,stricker2008fast,wang2010review}.
Statistical distributions of network states are essential to encode
information \cite{dan1998coding,vinje2000sparse,ohiorhenuan2010sparse,shemesh2013high,knill2004bayesian}.
For example, with statistical distributions of network states, experimental
studies show that rats can perform awake replays of remote experiences
in hippocampus \cite{karlsson2009awake}. Many works effectively characterize
the distribution of $2^{n}$ network states for $n$ binary-state
nodes in various systems,\emph{ }e.g., a network of $\sim100$ neurons
\cite{ganmor2011sparse}, with a\emph{ low-order} \emph{maximum entropy
principle} (MEP) analysis \cite{schneidman2006weak,shlens2006structure,tang2008maximum,marre2009prediction,bury2012statistical,watanabe2013pairwise,Barreiro2014microcircuits,martin2016pairwise}---a
method with few (far less than $2^{n}$) non-zero effective interactions
(see a precise definition in Eq. (\ref{eq:MaxEntModel})) constrained
by low-order statistics. We can then regard those effective interactions
as a sparse coding of the information that encoded in the state distribution.
To understand coding schemes of network systems, it is important,
however, yet to understand what leads to the sparsity of effective
interactions. In this work, we would mainly use neural networks as
examples for illustration, while our results apply to general binary-state
networks.

Estimated by dynamical data of a network system, effective interactions
reflect a dynamical structure of the network. This dynamical structure
has been used to study the functional connectivity of networks \cite{ganmor2011architecture,watanabe2013pairwise}.
For example, experimental studies show that the second-order effective
interaction map of the retina is sparse and dominated by local overlapping
effective interaction modules \cite{ganmor2011architecture}. Network
dynamical structure often closely relates to the underlying coupling
structure \cite{zhou2013causal}. For example, when the input of each
node is independent to others, i) high-order ($\geq2$) effective
interactions are zero in a network of no connections, ii) high-order
effective interactions are large in a dense and strong connected excitatory
network. To efficiently encode information, a realistic system often
incorporates a coupling structure with certain features \cite{newman2003structure,bullmore2009complex},
e.g., sparsity, small-world, or scale-free. However, it is still unclear
how the coupling structure affects the dynamical structure of effective
interactions.

In this letter, we consider a general class of pulse-coupled networks.
The state of each node is binary-state, i.e., active when the node
sends pulses to its child nodes, otherwise, silent. We observed a
\emph{Fact} that leads to an explicit relation---which is independent
of node dynamics---between the coupling structure and the number
of non-zero effective interactions in the full-order MEP analysis
(constrained by all moments). We examine our observed Fact by numerical
simulations. Through our analysis, we can estimate an upper bound
of the number of non-zero effective interactions for a given coupling
structure when the external input of each node is independent with
each other. Our results show that a sparse network could lead to a
lot of vanishing high-order effective interactions. For illustration,
we estimate the number of non-zero effective interactions for each
order in a network with\emph{ Erdos-Renyi} connection structure, in
which our estimation is much smaller than $C_{n}^{k}$, the number
of all possible $k$th-order effective interactions. Our results establish
a connection between the dynamical structure and the network coupling
structure. This connection provides an insight into how a sparse coupling
structure can lead to a sparse coding scheme.

In the following analysis, we use binary vector $V(l)=(\sigma_{1},\cdots,\sigma_{n})\in\{0,1\}^{n}$
to represent the state of $n$ nodes within the sampling time bin
labeled by $l$. To obtain correlations up to the $m$th-order requires
to evaluate all $\left\langle \sigma_{i_{1}}\cdots\sigma_{i_{M}}\right\rangle _{E}$,
where $1\leq i_{1}<i_{2}<\cdots<i_{M}\leq n$, $1\leq M\leq m$, and
$\left\langle \cdot\right\rangle _{E}$ is defined by$\left\langle g(l)\right\rangle _{E}=\sum_{l=1}^{N_{T}}g(l)/N_{T}$
for any function $g(l)$ and $N_{T}$ is the total number of sampling
time bins in the recording. The $m$th-order MEP analysis is to find
the desired probability distribution $P(V)$ for $n$ nodes by maximizing
the entropy $S\equiv-\sum_{V}P(V)\log P(V)$ subject to correlations
up to the $m$th-order ($m\leq n$). Then, the unique distribution
can be solved as 
\begin{equation}
P_{m}(V)=\frac{1}{Z}\exp(\sum_{k=1}^{m}\sum_{i_{1}<\cdots<i_{k}}^{n}J_{i_{1}\cdots i_{k}}\sigma_{i_{1}}\cdots\sigma_{i_{k}}),\label{eq:MaxEntModel}
\end{equation}
where, following the terminology of statistical physics, we call $J_{i_{1}\cdots i_{k}}$
a $k$th-order effective interaction ($1\leq k\leq m$), the partition
function $Z$ is the normalization factor. Eq. (\ref{eq:MaxEntModel})
is referred to as the $m$th-order MEP distribution.

\emph{First}, we discuss the relationship between effective interactions
and the statistical distribution of network states. By taking logarithm
of both sides of Eq. (\ref{eq:MaxEntModel}) for $P_{n}(V)$, we can
get a set linear equations of all-order effective interactions for
all states $V$. Since $P_{n}$ is the same as the experimental observed
distribution \cite{amari2001information}, we can obtain the effective
interactions in $P_{n}$ in terms of the experimental observed distribution
\cite{xu2016dynamical}. For example, $n=3$, we can obtain $J_{1}=\log(P_{100}/P_{000})$
and $J_{12}=\log(P_{110}/P_{010})-J_{1}$, where $P_{\sigma_{1}\sigma_{2}\sigma_{3}}$
represents the probability of the network state $(\sigma_{1},\sigma_{2},\sigma_{3})$.
By applying $P(\sigma_{1},\sigma_{2},\sigma_{3})=P(\sigma_{1}|\sigma_{2},\sigma_{3})P(\sigma_{2},\sigma_{3})$,
we have $J_{1}=\log\frac{P(\sigma_{1}=1|\sigma_{2}=0,\sigma_{3}=0)}{P(\sigma_{1}=0|\sigma_{2}=0,\sigma_{3}=0)}$
and $J_{12}=\log\frac{P(\sigma_{1}=1|\sigma_{2}=1,\sigma_{3}=0)}{P(\sigma_{1}=0|\sigma_{2}=1,\sigma_{3}=0)}-J_{1}\triangleq J_{1}^{1}-J_{1}$.
Our earlier study has shown a recursive structure among effective
interactions, that is, the $(k+1)$st-order effective interaction
$J_{123\dots(k+1)}$ can be obtained as follows \cite{xu2016dynamical}:
First, we switch the state of the $(k+1)$st node in $J_{123\dots k}$
from silent to active to obtain a new term $J_{123\dots k}^{1}$,
e.g., from $J_{1}$ to $J_{1}^{1}$; Then, we subtract $J_{123\dots k}$
from the new term to obtain $J_{123\dots(k+1)}$, i.e., 
\begin{equation}
J_{123\dots(k+1)}=J_{123\dots k}^{1}-J_{123\dots k}.\label{eq:recursive}
\end{equation}
Without lost of generality, we randomly select two nodes labeled by
$1$ and $2$. By the recursive relation, any $k$th-order effective
interaction that includes node 1 and 2 can be expressed as the summation
of terms with the following basic form 
\begin{align}
J_{12}^{b}(\sigma_{3},\cdots,\sigma_{n}) & =\log\frac{P(\sigma_{1}=1|\sigma_{2}=1,\sigma_{3},\cdots,\sigma_{n})}{P(\sigma_{1}=0|\sigma_{2}=1,\sigma_{3},\cdots,\sigma_{n})}\nonumber \\
 & \quad-\log\frac{P(\sigma_{1}=1|\sigma_{2}=0,\sigma_{3},\cdots,\sigma_{n})}{P(\sigma_{1}=0|\sigma_{2}=0,\sigma_{3},\cdots,\sigma_{n})}.\label{eq:J12Cond-1}
\end{align}
For example, $J_{123}=J_{12}^{b}(1,0,\cdots,0)-J_{12}^{b}(0,0,\cdots,0)$
and $J_{1234}=[J_{12}^{b}(1,1,0,\cdots,0)-J_{12}^{b}(0,1,0,\cdots,0)]-J_{123}$.
We can observe that if nodes $1$ and $2$ are independent conditioned
on all other nodes, i.e., $P(\sigma_{1}|\sigma_{2}=1,\sigma_{3},\cdots,\sigma_{n})=P(\sigma_{1}|\sigma_{2}=0,\sigma_{3},\cdots,\sigma_{n})$,
any effective interaction containing these two nodes is zero.

\emph{Next}, we would show what kind of coupling structure could entail
the conditional independence of two nodes. Here, we define some notations.
In any sampling time bin $[0,\Delta)$ with state $V=(\sigma_{1},\cdots,\sigma_{n})$,
$\forall t\in[0,\Delta)$, we denote $I_{i,t}$ as node $i$'s input
from the outside of the network, denote $w_{ij}(t)$ as the input
from the node $i$ to node $j$, denote ${\rm C}(i)$ as the set of
all child notes of node $i$, denote $U_{i}={\rm C}(i)\cup\{i\}$,
denote $P(e)$ as the probability of event $e$, denote $U_{0}=\{1,2,\cdots,n\}$. 
\begin{fact*}
For $n$ pulse-coupled nodes with binary-state dynamics on a network
with a coupling structure $G_{0}$, in any sampling time bin $[0,\Delta)$,
$\forall t\in[0,\Delta)$, $\forall i_{1},j_{1}\in U_{0}$, we assume
that: (a) the external inputs of each node are independent to others,
i.e., $P(I_{i_{1},t},I_{j_{1},t})=P(I_{i_{1},t})P(I_{j_{1},t})$;
(b) whether a parent node sends spikes to its child nodes only depends
on its state, i.e., $P(w_{i_{1}j_{1}}(t),V)=W(\sigma_{i_{1}},i_{1},j_{1},t)$,
where $W(\cdot,\cdot,\cdot,\cdot)$ is a real function. $\forall i,j\in U_{0}$,
if they neither are connected nor share any common child node, i.e.,
$U_{i}\cap U_{j}=\phi$, then, node $i$ and $j$ are independent
conditioned on the state of all other nodes, i.e., 
\begin{equation}
P(\sigma_{i},\sigma_{j}|H)=P(\sigma_{i}|H)P(\sigma_{j}|H),\label{eq:CondIndProb-1}
\end{equation}
where $H$ is a possible state of nodes in $U_{0}\backslash\{i,j\}$. 
\end{fact*}
We justify our two assumptions as follows. To avoid the influence
of correlation in external inputs when we are studying the relation
between the dynamical structure and the coupling structure, we assume
that the external input of each node is independent to others, i.e.,
assumption (a). The second assumption implicates a Markov-like property;
that is, for a connected pair of pulse-coupled nodes in an equilibrium
state, the pulse from the parent node to the child node only depends
on the state of the parent node but is independent of inputs imposed
on the parent node. For example, in neural networks, a neuron sends
out spikes only when this neuron is active, regardless of what inputs
are imposed on the neuron.

The argument for the conclusion in Eq. (\ref{eq:CondIndProb-1}) is
as follows. By assumption (a), node $i$ and node $j$ can be dependent
only through the coupling structure $G_{0}$. When we are considering
how node $i$ and node $j$ affect each other by changing their states
through the coupling structure $G_{0}$, we can consider a simplified
coupling structure, $G_{1}$, which ignores those connections that
are independent of states of node $i$ and node $j$, i.e., $\sigma_{i}$
and $\sigma_{j}$. $\forall k\in U_{o}\backslash\{i,j\}$, i.e., any
other node $k$, its state $\sigma_{k}$ is fixed when we are considering
the conditional probability in Eq. (\ref{eq:CondIndProb-1}). By assumption
(b), for node $k$'s any child node $l$, the input from node $k$
to node $l$ is independent of $\sigma_{i}$ and $\sigma_{j}$. Thus,
the connections started from those nodes in $U_{o}\backslash\{i,j\}$
are fixed for different states of $\sigma_{i}$ and $\sigma_{j}$.
Therefore, $G_{1}$ is a simplified coupling structure that only keeps
those connections originated from node $i$ and node $j$ in $G_{0}$.
In $G_{1}$, any connection only exists in either sub-network $U_{i}$
or sub-network $U_{j}$. Under the condition $U_{i}\cap U_{j}=\phi$,\emph{
}i.e., they neither are connected nor share any common child node,
sub-network $U_{i}$ and sub-network $U_{j}$ are two isolated sub-networks.
$\sigma_{i}$ and $\sigma_{j}$ cannot affect each other by changing
their states through the coupling structure $G_{1}$, that is, node
$i$ and $j$ are independent conditioned on the states of all other
nodes.

\begin{figure}
\begin{centering}
\subfloat[ ]{\begin{centering}
\includegraphics[scale=0.32]{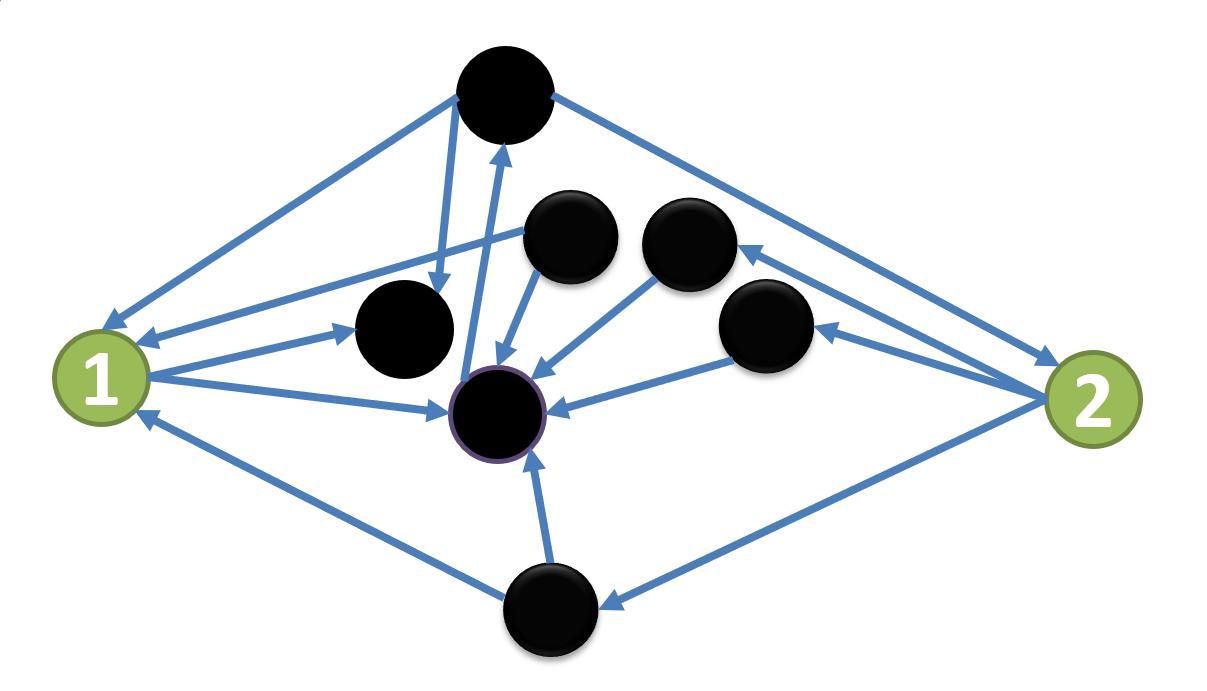} 
\par\end{centering}
}\subfloat[ ]{\begin{centering}
\includegraphics[scale=0.32]{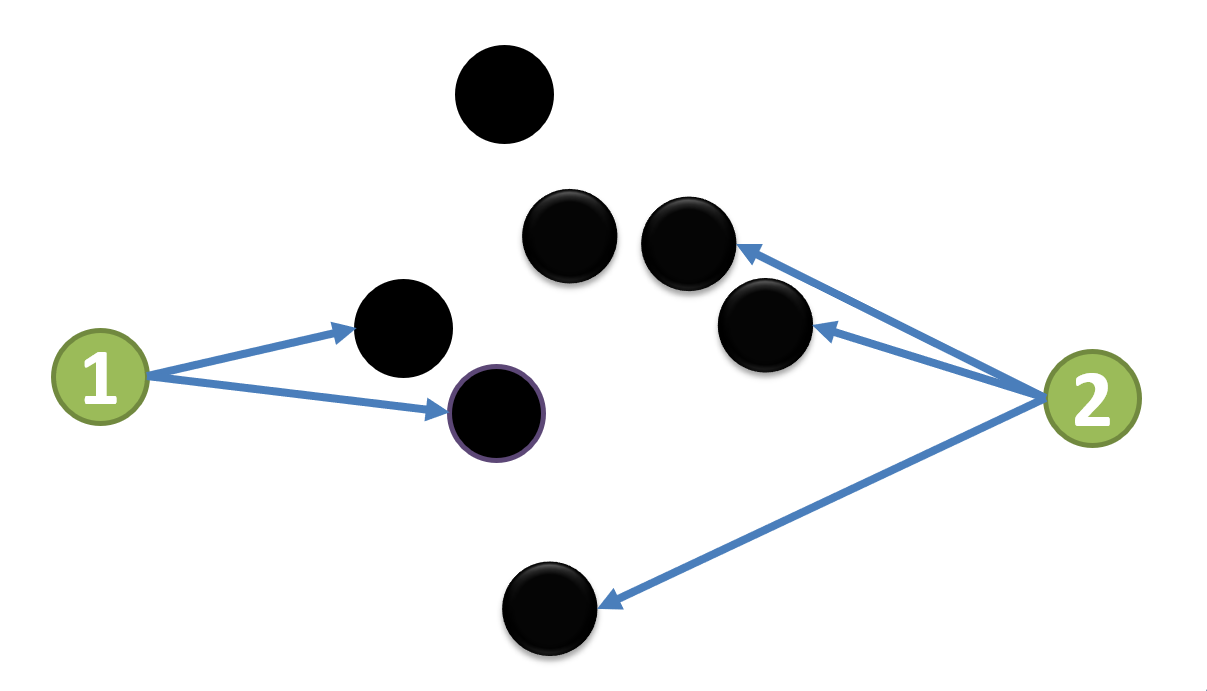} 
\par\end{centering}
}
\par\end{centering}
\caption{\textbf{Structure v.s. Simplified Structure.} \label{fig:Structure}}
\end{figure}

Fig.\ref{fig:Structure} displays an example to illustrate our observed
Fact. The coupling structure $G_{0}$ is shown in Fig.\ref{fig:Structure}a.
We focus on node $1$ and node $2$, where they neither are connected
nor share any child node. When the state of other nodes (black) are
fixed, all outputs from black nodes can be ignored in the simplified
coupling structure $G_{1}$, as shown in Fig.\ref{fig:Structure}b.
Node $1$ and node $2$ respectively belong to two separate sub-networks.
Therefore, nodes $1$ and node $2$ are independent conditioned on
the state of all other nodes.

Based on the recursive structure of effective interactions and the
observed Fact, we reach the following conclusion: with the two assumptions
in the observed Fact, for a group of nodes $\{i_{1},i_{2},\cdots,i_{k}\}$,
if there exists at least one pair of nodes that neither are connected
nor share any child node, effective interaction $J_{i_{1},i_{2},\cdots,i_{k}}$
is zero.

In the system we would use to examine our conclusion is an integrate-and-fire
(I\&F) network, a general pulse-coupled network, with both excitatory
and inhibitory nodes \cite{zhou2013causal}. For the $i$th node,
the dynamics of its state variable $x_{i}$ with time scales $\tau$
is governed by 
\begin{equation}
\dot{x}_{i}=-\frac{x_{i}}{\tau}-(g_{i}^{{\rm bg}}+g^{{\rm ex}})(x_{i}-x_{{\rm ex}})-g_{i}^{{\rm in}}(x_{i}-x_{{\rm in}}),\label{eq:IFModel}
\end{equation}
where $x_{{\rm ex}}$ and $x_{{\rm in}}$ are the reversal values
of excitation (ex) and inhibition (in), respectively. $g_{i}^{{\rm bg}}=f\sum_{k}\mathbb{H}(t-T_{i,k}^{F})\exp[-(t-T_{i,k}^{F})/\sigma^{{\rm ex}}]$
is the background input with magnitude $f$ and time scale $\sigma^{{\rm ex}}$,
$T_{i,k}^{F}$ is a Poisson process with rate $\mu$, $\mathbb{H}(\cdot)$
is the Heaviside function, $g_{i}^{{\rm ex}}=\sum_{j}\sum_{k}S_{ij}^{{\rm ex}}\mathbb{H}(t-T_{j,k}^{{\rm ex}})\exp[-(t-T_{j,k}^{{\rm ex}})/\sigma^{{\rm ex}}]$
is the excitatory pulse effective interaction from other $j$th excitatory
nodes, and $g_{i}^{{\rm in}}=\sum_{j}\sum_{k}S_{ij}^{{\rm in}}\mathbb{H}(t-T_{j,k}^{{\rm in}})\exp[-(t-T_{j,k}^{{\rm in}})/\sigma^{{\rm in}}]$
is the inhibitory pulse effective interaction from other $j$th inhibitory
nodes. The $j$th excitatory (inhibitory) node $x_{j}$ evolves continuously
according to Eq. (\ref{eq:IFModel}) until it reaches a firing threshold
$x_{{\rm th}}$. That moment in time is referred to as a firing event
(say, the $k$th spike) and denoted by $T_{j,k}^{{\rm ex}}$ ($T_{j,k}^{{\rm in}}).$
Then, $x_{j}$ is reset to the reset value $x_{r}$ ($x_{{\rm in}}<x_{r}<x_{{\rm th}}<x_{\text{{\rm ex}}}$)
and held $x_{r}$ for an absolute refractory period of $\tau_{{\rm ref}}.$
Each spike emerging from the $j$th excitatory (inhibitory) node causes
an instantaneous increase $S_{ij}^{{\rm ex}}$ ($S_{ij}^{{\rm in}}$)
in $g_{i}^{{\rm ex}}$ ($g_{i}^{{\rm in}}$), where $S_{ij}^{{\rm ex}}$
and $S_{ij}^{{\rm in}}$ are the excitatory and inhibitory coupling
strengths, respectively. The model (\ref{eq:IFModel}) describes a
general class of physical networks \cite{mirollo1990synchronization,gerstner2002spiking,cai2005architectural,wang2010review,zhou2013causal}.%

\begin{figure*}
\begin{centering}
\subfloat[ ]{\begin{centering}
\includegraphics[scale=0.46]{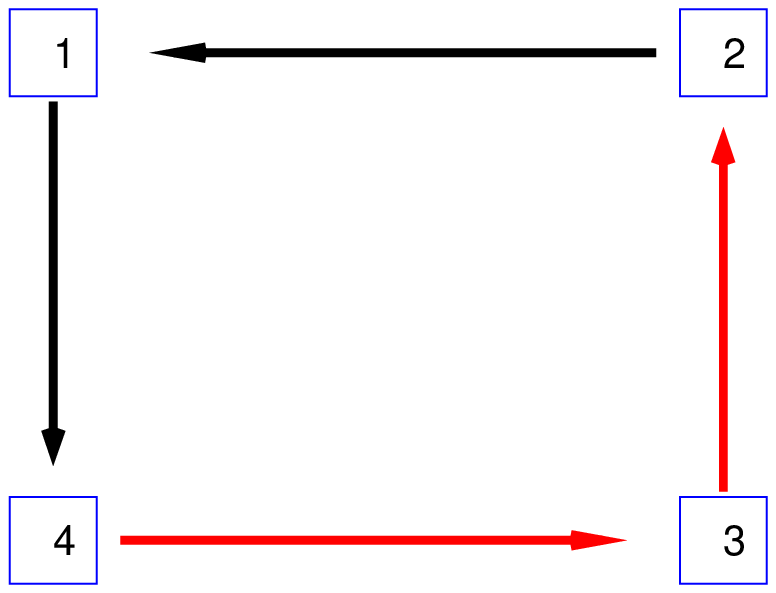} 
\par\end{centering}
}\subfloat[ ]{\begin{centering}
\includegraphics[scale=0.26]{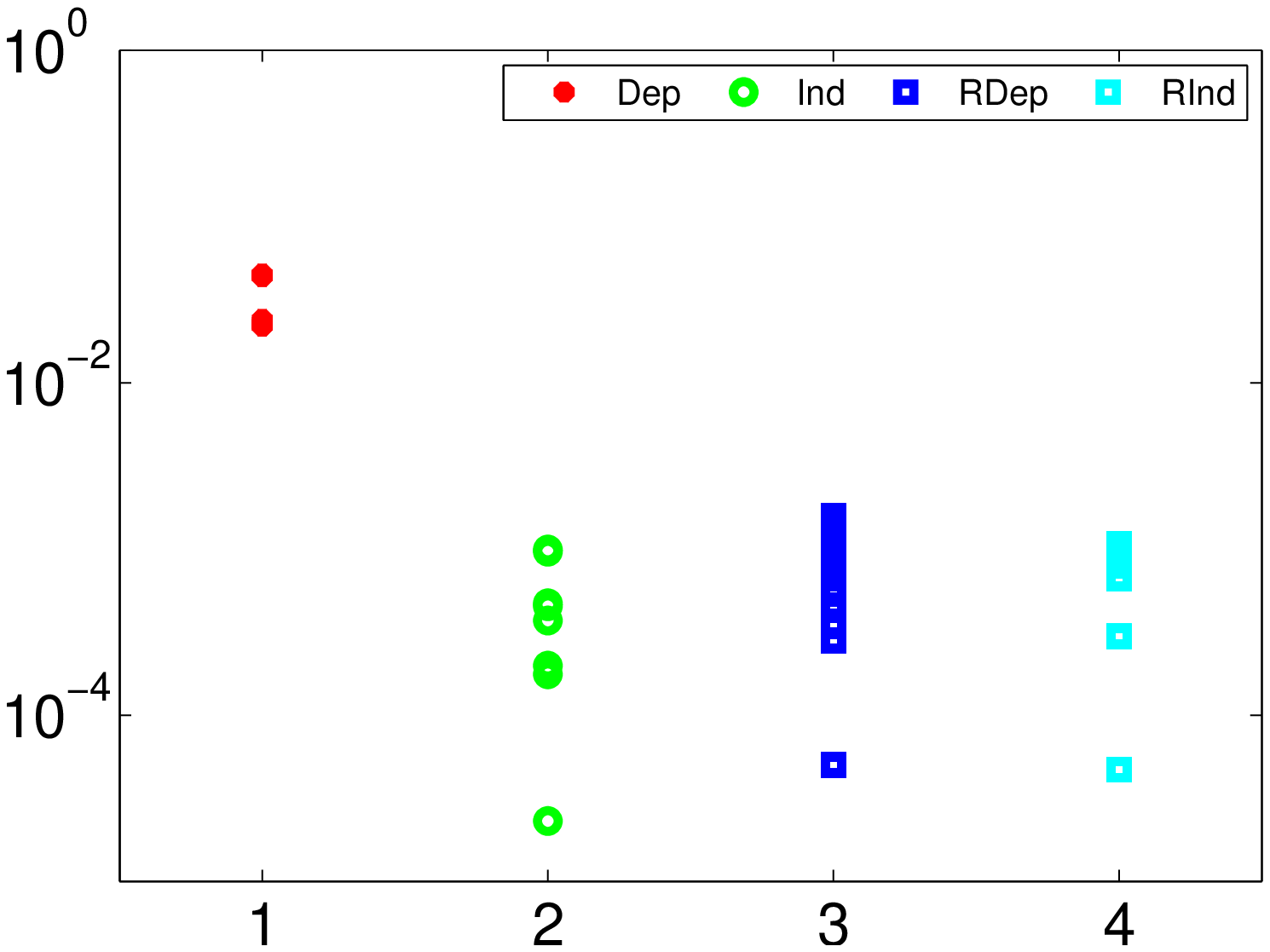} 
\par\end{centering}
}\subfloat[ ]{\begin{centering}
\includegraphics[scale=0.23]{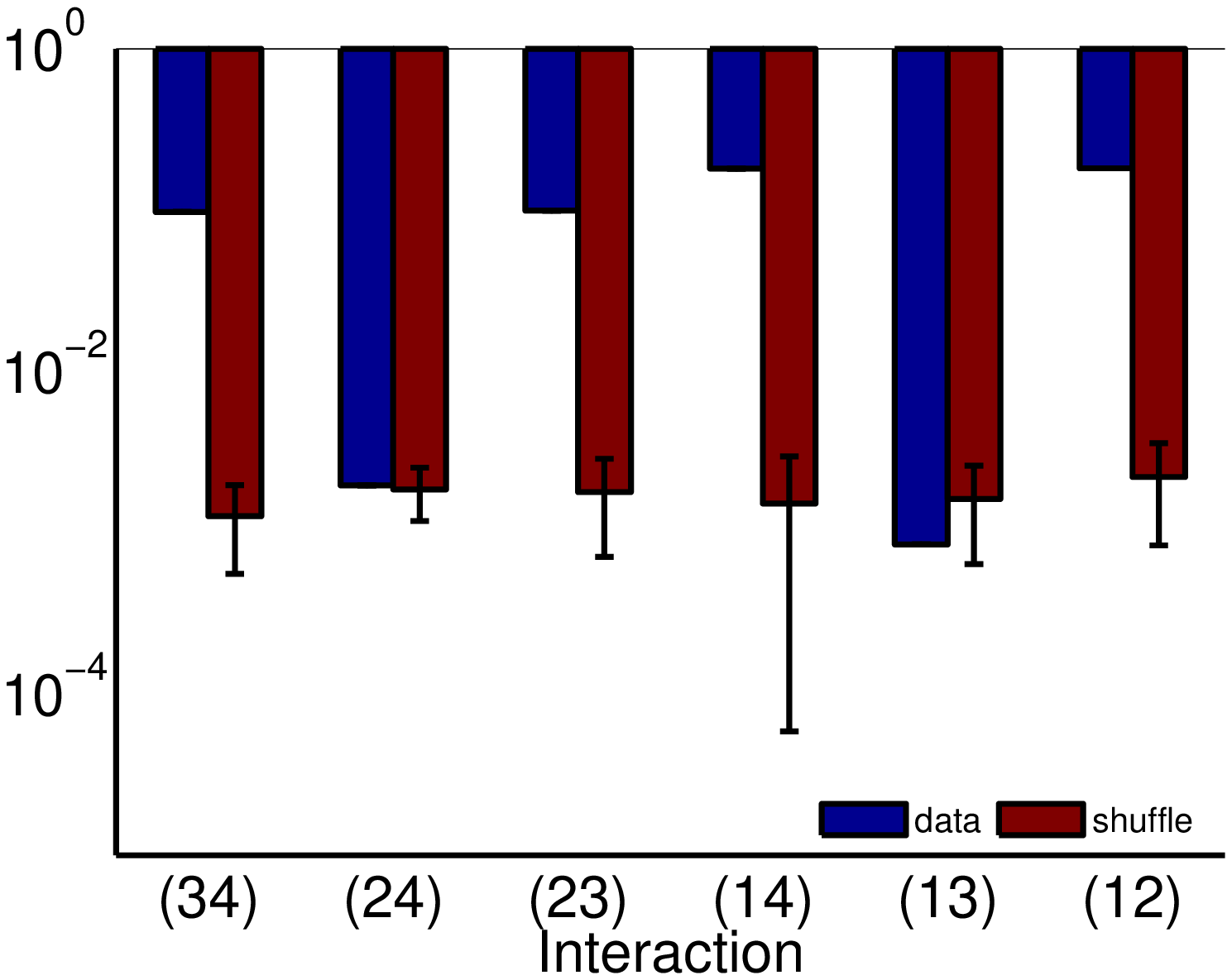} 
\par\end{centering}
}\subfloat[ ]{\begin{centering}
\includegraphics[scale=0.23]{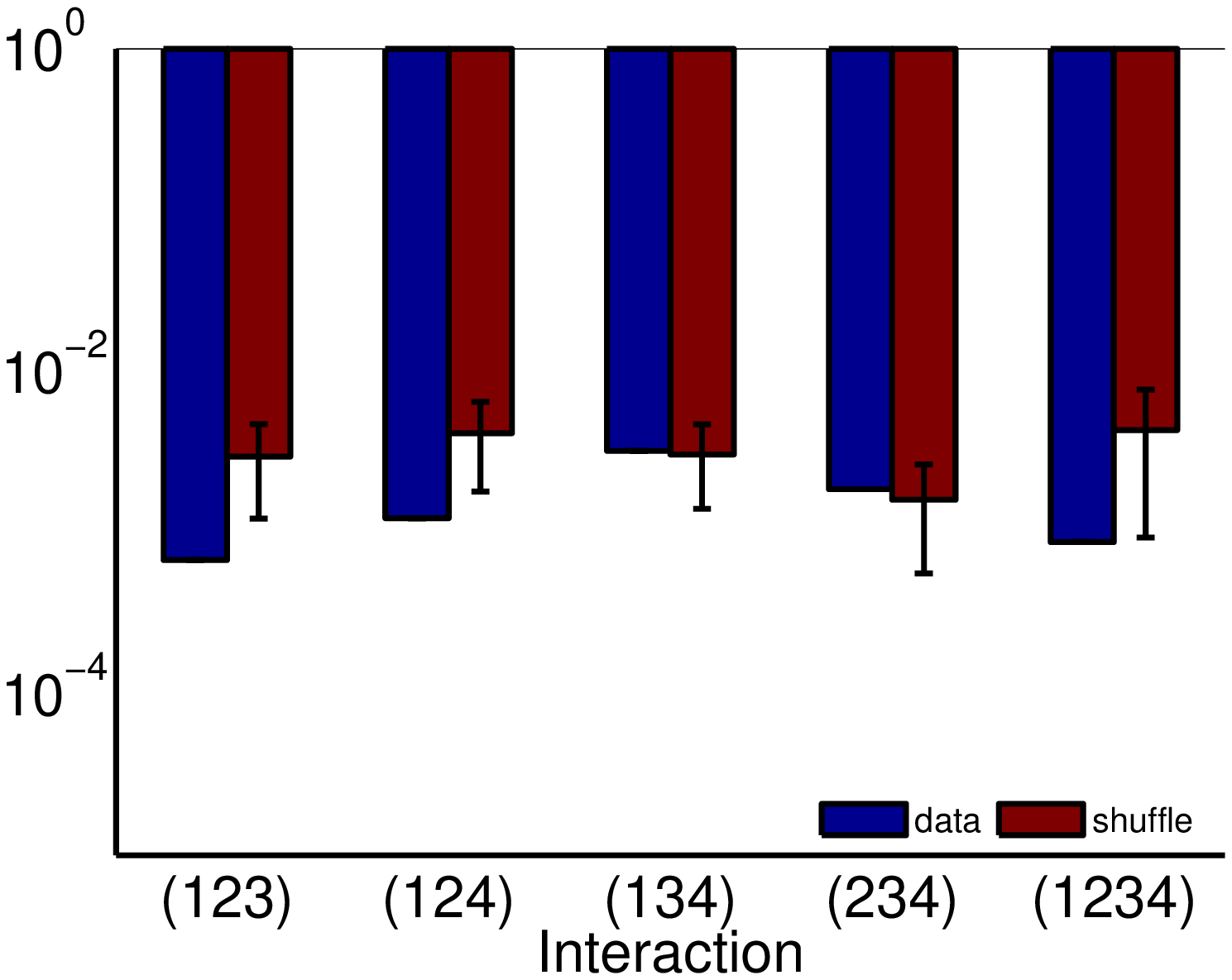} 
\par\end{centering}
}
\par\end{centering}
\begin{centering}
\subfloat[ ]{\begin{centering}
\includegraphics[scale=0.46]{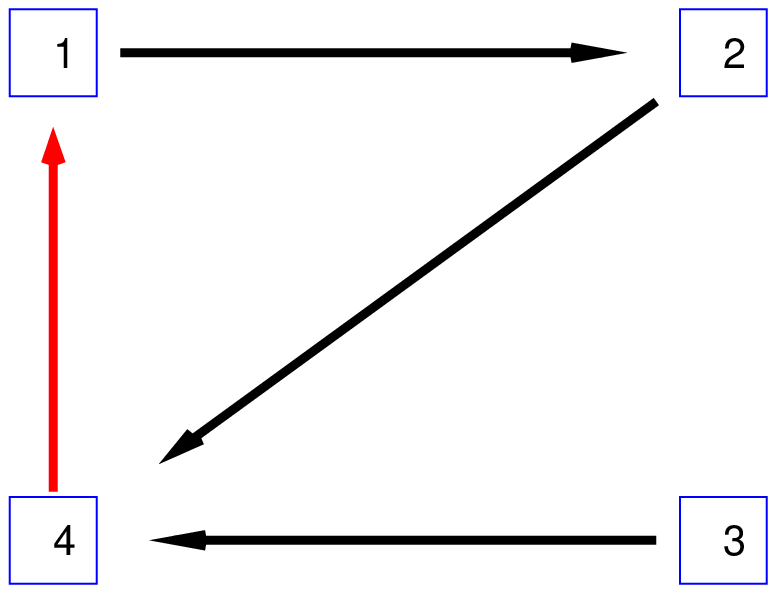} 
\par\end{centering}
}\subfloat[ ]{\begin{centering}
\includegraphics[scale=0.26]{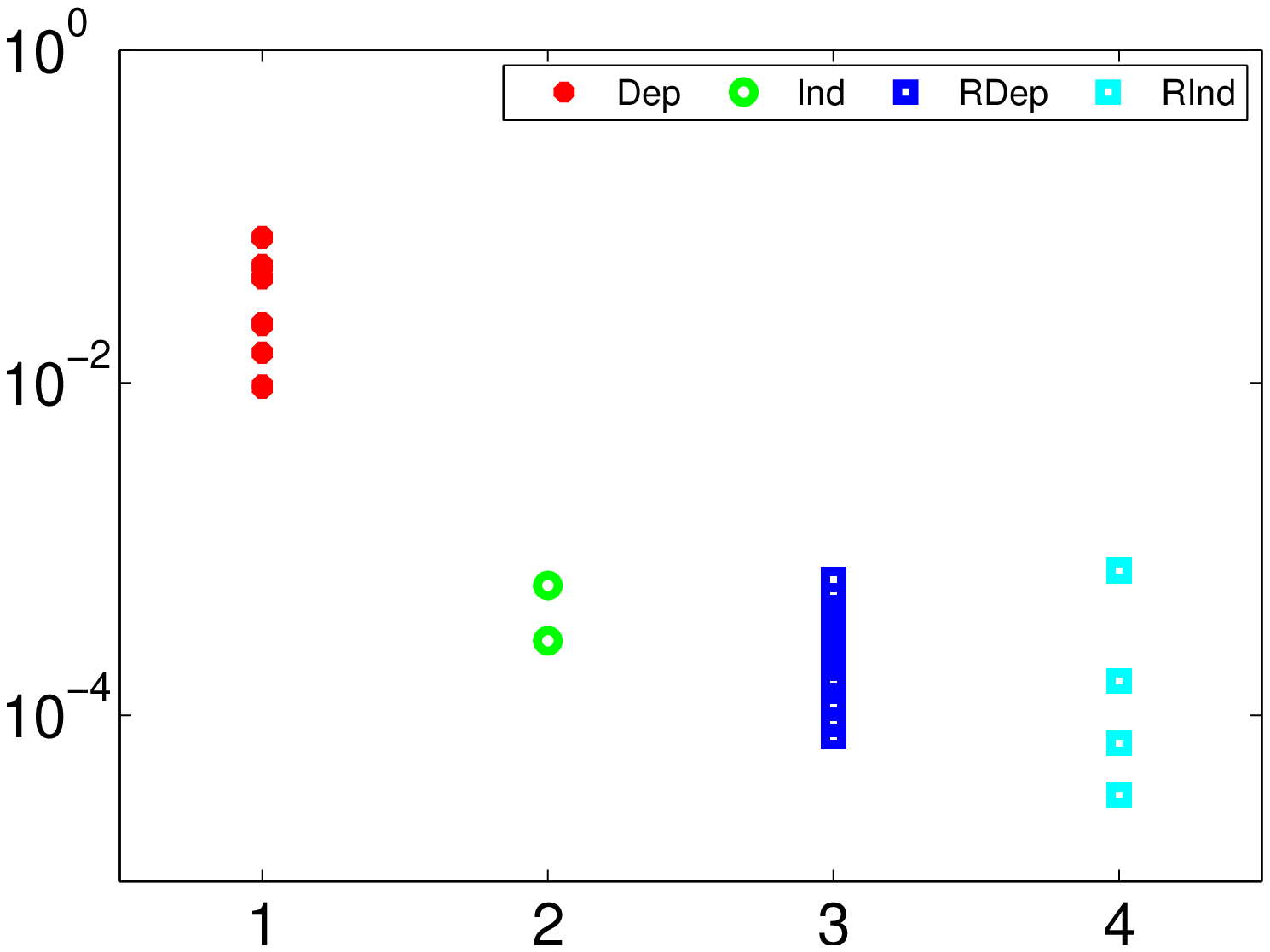} 
\par\end{centering}
}\subfloat[ ]{\begin{centering}
\includegraphics[scale=0.23]{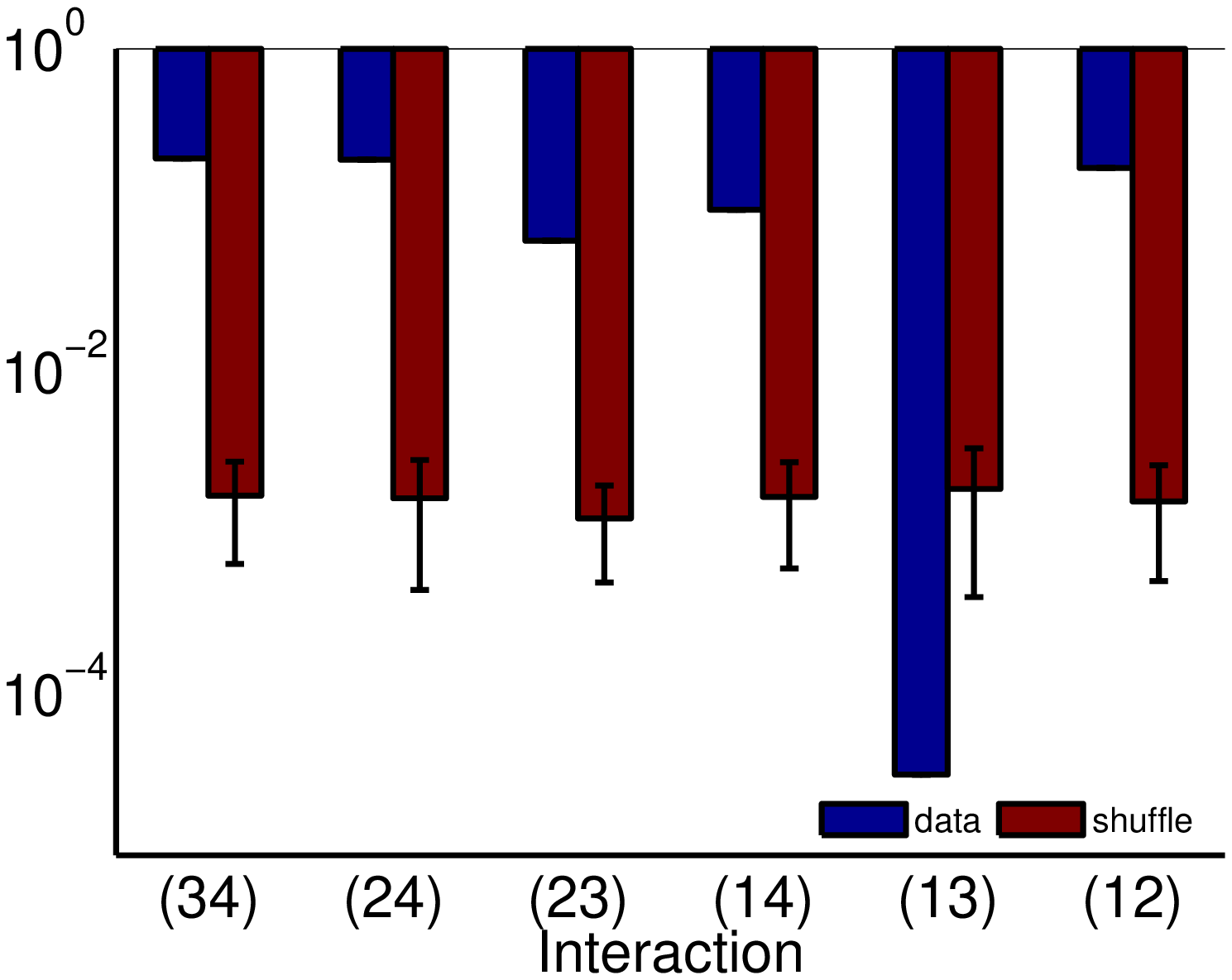} 
\par\end{centering}
}\subfloat[ ]{\begin{centering}
\includegraphics[scale=0.23]{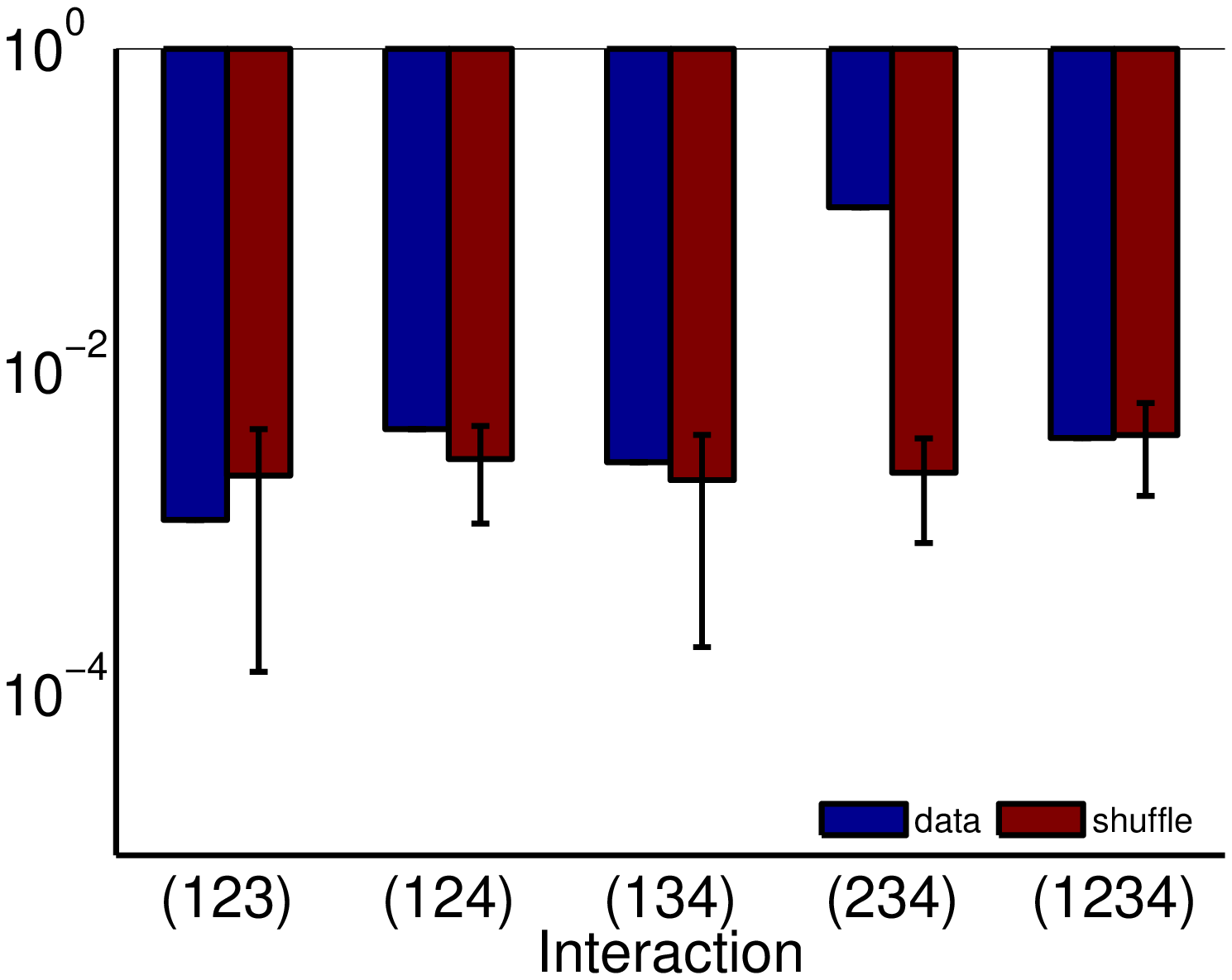} 
\par\end{centering}
}
\par\end{centering}
\caption{\textbf{Structure v.s. effective interactions of I\&F networks .}
Each row shows a numerical case. In the first column, black arrows
and red arrows represents excitatory and inhibitory connections, respectively.
In the second column, red and green dots are the strengths of $\Delta_{ij}(H)$
of dependent and independent pairs, respectively. Blue dots and cyan
dots are the strengths of $\Delta_{ij}(H)$ of dependent and independent
pairs from ten shuffled spike trains, respectively. The third and
fourth columns display absolute effective interaction strengths (blue
bars). The corresponding node indexes for each effective interaction
are shown in the abscissa. The mean and standard deviation of absolute
strengths of each effective interaction of ten shuffled spike trains
are also displayed by garnet bars. The simulation time for each network
is $\unit[1.2\times10^{8}]{ms}$. The time bin size for analysis is
$\unit[10]{ms}$ \cite{shlens2006structure,tang2008maximum}. Independent
Poisson inputs for each network are $\mu=0.\unit[1]{ms^{-1}}$ and
$f=\unit[0.1]{ms^{-1}}$. The firing rate of each node is about $\unit[50]{Hz}$.
Parameters are chosen \cite{gerstner2002spiking} as $x_{{\rm ex}}=14/3$,
$x_{{\rm in}}=-2/3$, $\sigma^{{\rm ex}}=\unit[2]{ms}$, $\sigma^{{\rm in}}=\unit[5]{ms}$,
$\tau=\unit[20]{ms}$, $x_{{\rm th}}=1$, $x_{r}=0$, and $\tau_{{\rm ref}}=\unit[2]{ms}$,
$S_{ij}^{{\rm ex}}=S_{ij}^{{\rm in}}=0.02$.\label{fig:IF4Cases}}
\end{figure*}

The first example, two excitatory and two inhibitory I\&F nodes form
a ring coupling structure (Fig.\ref{fig:IF4Cases}a). For any pair
of nodes, say, node $i$ and $j$, we compute $\Delta_{ij}(H)=|P(\sigma_{i}=1|\sigma_{j}=1,H)-P(\sigma_{i}=1|\sigma_{j}=0,H)|$,
where $H$ is one state of other two nodes. By our observed Fact,
the conditional independent pairs are $(\text{{\rm neuron }}1,\text{{\rm neuron }}3)$
and $(\text{{\rm neuron }}2,\text{{\rm neuron }}4)$, and other pairs
are categorized as dependent pairs. In Fig.\ref{fig:IF4Cases}b, the
strengths of $\Delta_{ij}(H)$ of independent pairs (green) are almost
two orders of magnitude smaller than those of dependent pairs (red).
We then shuffle spike trains of each node. We similarly compute $\Delta_{ij}(H)$
for $10$ different shuffled data. Blue dots and cyan dots in Fig.\ref{fig:IF4Cases}b
are results of all shuffled data of dependent pairs and independent
pairs, respectively. The strength of $\Delta_{ij}(H)$ of independent
pairs (green)---computed from the observed data---are within the
statistical error of shuffled data. We then solve effective interactions
in the full-order MEP analysis $P_{n}$ for this ring network. As
shown in Fig.\ref{fig:IF4Cases}c, the effective interaction strengths
of independent pairs ($J_{24}$ and $J_{13})$ are within the statistical
error of shuffled results (red). Since every high-order ($\geq3$)
effective interaction includes at least one independent pair of nodes,
as predicted, the strengths of all high-order effective interactions
are within the statistical error of shuffled results as shown in Fig.\ref{fig:IF4Cases}d.

The second example in the second row in Fig.\ref{fig:IF4Cases}, results
are similar that dependent pairs and independent pairs can be identified
through our observed Fact, and the strength of any effective interaction
that includes the independent pair of nodes (node 1 and node 3) is
within the statistical error of shuffled data. In this example, $J_{124}$
is very small, i.e., within the statistical error of shuffled results.
However, in our estimation by our conclusion, we do not categorized
$J_{124}$ to the class of zero-strength effective interactions. This
example indicates that we estimate an upper bound of the number of
non-zero effective interactions. For a network of all excitatory nodes
with the same coupling structure as the one in Fig.\ref{fig:Structure}e,
$J_{124}$ is significantly larger than zero (not shown). Since the
strength of high-order effective interactions is small, a very long
recording constraints us from examining $\Delta_{ij}(H)$ for a large
network.

\begin{figure}
\begin{centering}
\includegraphics[scale=0.34]{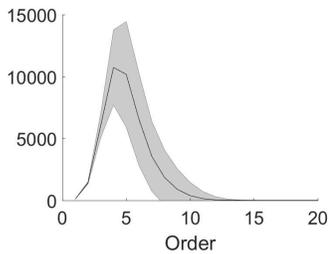} 
\par\end{centering}
\caption{\textbf{Network of Erdos-Renyi connection.} We randomly generate $1000$
networks of $100$ nodes with an \emph{Erdos-Renyi} connection. The
connection probability between two nodes is $0.05$. The number of
non-zero effective interaction v.s. effective interaction order. The
mean and standard deviation are respectively shown by the black line
and shaded area. \label{fig:ERJnum}}
\end{figure}

Base on the relation between the coupling structure and effective
interactions, the number of non-zero high-order effective interactions
can be small in a sparse connected network compared with $C_{n}^{k}$,
which is the number of all possible $k$th-order interactions. For
example, we estimate the number of each-order non-zero effective interactions
in a network with an \emph{Erdos-Renyi} connection structure. We randomly
generate $1000$ networks of $100$ nodes with an \emph{Erdos-Renyi}
connection. The connection probability between two nodes is $0.05$.
As shown in Fig.\ref{fig:ERJnum}, the number of non-zero $k$th-order
($k>1$) effective interactions is much smaller than $C_{100}^{k}$
(too large to be shown). The number of high-order effective interactions
(order higher than $11$th) almost vanishes (order higher than $20$th
not shown).

In summary, we have established a relation between effective interactions
in MEP analysis and the coupling structure of pulse-coupled networks
to understand how a sparse coupling structure could lead to a sparse
coding by effective interactions. This relation quantitatively displays
how the dynamical structure closely relates to the coupling structure.

Even though high-order effective interactions are often much smaller
compared with low-order ones \cite{xu2016dynamical}, it is still
unclear why small high-order effective interactions do not accumulate
to have a significant effect in a large network \cite{shlens2009structure,ganmor2011sparse}.
For example, MEP distribution with a sparse low-order effective interactions---non-zero
effective interactions are sparse and vanish when the order is high
than the eighth-order---can well capture the state distribution of
$99$ ganglion cells in the salamander retina responding to a natural
movie clip or natural pixel \cite{ganmor2011sparse}. In this study,
we show that a large amount of effective interactions vanish in a
sparse coupling structure; thus, rationalizing the absence of the
accumulation of high-order interactions for a large network.

Finally, we point out that some important issues remain to be elucidated
in the future. First, we have ignored correlations in external inputs
when estimating the number of non-zero effective interactions. Correlated
inputs can induce non-zero high-order effective interactions \cite{macke2011common}.
It is yet to consider how the statistics of inputs affect the sparsity
of effective interactions. Second, current algorithms for estimating
non-zero effective interactions (not limited to the second-order)
for a large network (e.g., $\sim100$ nodes) are very slow, e.g.,
Monte Carlo based methods \cite{shlens2009structure,nasser2013spatio}.
Our undergoing work is exploring a fast algorithm that exploits the
sparsity of effective interactions. We have seen an indication that
the algorithm can work well for an I\&F network with sparse coupling
structure; however, that work is yet to be fully verified to be conclusive.
\begin{acknowledgments}
The authors thank David W. McLaughlin for helpful discussions. This
work was supported by NSFC-11671259, NSFC-11722107, NSFC-91630208
and Shanghai Rising-Star Program-15QA1402600 (D.Z.); by NSF DMS-1009575
and NSFC-31571071 (D.C.); by Shanghai 14JC1403800, 15JC1400104, and
SJTU-UM Collaborative Research Program (D.C. and D.Z.); and by the
NYU Abu Dhabi Institute G1301 (Z.X., D.Z., and D.C.). 
\end{acknowledgments}

\subsection*{}

 \bibliographystyle{SIAM}
\bibliography{MERef}

\begin{thebibliography}{10}

\bibitem{amari2001information}
{\sc S.-I. Amari}, {\em Information geometry on hierarchy of probability
  distributions}, Information Theory, IEEE Transactions on, 47 (2001),
  pp.~1701--1711.

\bibitem{Barreiro2014microcircuits}
{\sc A.~K. Barreiro, J.~Gjorgjieva, F.~Rieke, and E.~Shea-Brown}, {\em When do
  microcircuits produce beyond-pairwise correlations?}, Frontiers in
  computational neuroscience, 8 (2014).

\bibitem{bullmore2009complex}
{\sc E.~Bullmore and O.~Sporns}, {\em Complex brain networks: graph theoretical
  analysis of structural and functional systems}, Nature Reviews Neuroscience,
  10 (2009), pp.~186--198.

\bibitem{bury2012statistical}
{\sc T.~Bury}, {\em Statistical pairwise interaction model of stock market},
  The European Physical Journal B, 86 (2013), p.~89.

\bibitem{cai2005architectural}
{\sc D.~Cai, A.~V. Rangan, and D.~W. McLaughlin}, {\em Architectural and
  synaptic mechanisms underlying coherent spontaneous activity in v1},
  Proceedings of the National Academy of Sciences of the United States of
  America, 102 (2005), pp.~5868--5873.

\bibitem{dan1998coding}
{\sc Y.~Dan, J.-M. Alonso, W.~M. Usrey, and R.~C. Reid}, {\em Coding of visual
  information by precisely correlated spikes in the lateral geniculate
  nucleus}, Nature neuroscience, 1 (1998), pp.~501--507.

\bibitem{ganmor2011architecture}
{\sc E.~Ganmor, R.~Segev, and E.~Schneidman}, {\em The architecture of
  functional interaction networks in the retina}, The journal of neuroscience,
  31 (2011), pp.~3044--3054.

\bibitem{ganmor2011sparse}
{\sc E.~Ganmor, R.~Segev, and E.~Schneidman}, {\em Sparse low-order interaction
  network underlies a highly correlated and learnable neural population code},
  Proceedings of the National Academy of Sciences, 108 (2011), pp.~9679--9684.

\bibitem{gerstner2002spiking}
{\sc W.~Gerstner and W.~M. Kistler}, {\em Spiking neuron models: Single
  neurons, populations, plasticity}, Cambridge university press, 2002.

\bibitem{karlsson2009awake}
{\sc M.~P. Karlsson and L.~M. Frank}, {\em Awake replay of remote experiences
  in the hippocampus}, Nature neuroscience, 12 (2009), p.~913.

\bibitem{knill2004bayesian}
{\sc D.~C. Knill and A.~Pouget}, {\em The bayesian brain: the role of
  uncertainty in neural coding and computation}, TRENDS in Neurosciences, 27
  (2004), pp.~712--719.

\bibitem{macke2011common}
{\sc J.~H. Macke, M.~Opper, and M.~Bethge}, {\em Common input explains
  higher-order correlations and entropy in a simple model of neural population
  activity}, Physical Review Letters, 106 (2011), p.~208102.

\bibitem{marre2009prediction}
{\sc O.~Marre, S.~El~Boustani, Y.~Fr{\'e}gnac, and A.~Destexhe}, {\em
  Prediction of spatiotemporal patterns of neural activity from pairwise
  correlations}, Physical review letters, 102 (2009), p.~138101.

\bibitem{martin2016pairwise}
{\sc E.~A. Martin, J.~Hlinka, and J.~Davidsen}, {\em Pairwise network
  information and nonlinear correlations}, Physical Review E, 94 (2016),
  p.~040301.

\bibitem{mirollo1990synchronization}
{\sc R.~E. Mirollo and S.~H. Strogatz}, {\em Synchronization of pulse-coupled
  biological oscillators}, SIAM Journal on Applied Mathematics, 50 (1990),
  pp.~1645--1662.

\bibitem{nasser2013spatio}
{\sc H.~Nasser, O.~Marre, and B.~Cessac}, {\em Spatio-temporal spike train
  analysis for large scale networks using the maximum entropy principle and
  monte carlo method}, Journal of Statistical Mechanics: Theory and Experiment,
  2013 (2013), p.~P03006.

\bibitem{newman2003structure}
{\sc M.~E. Newman}, {\em The structure and function of complex networks}, SIAM
  review, 45 (2003), pp.~167--256.

\bibitem{ohiorhenuan2010sparse}
{\sc I.~E. Ohiorhenuan, F.~Mechler, K.~P. Purpura, A.~M. Schmid, Q.~Hu, and
  J.~D. Victor}, {\em Sparse coding and high-order correlations in fine-scale
  cortical networks}, Nature, 466 (2010), pp.~617--621.

\bibitem{schneidman2006weak}
{\sc E.~Schneidman, M.~J. Berry, R.~Segev, and W.~Bialek}, {\em Weak pairwise
  correlations imply strongly correlated network states in a neural
  population}, Nature, 440 (2006), pp.~1007--1012.

\bibitem{shemesh2013high}
{\sc Y.~Shemesh, Y.~Sztainberg, O.~Forkosh, T.~Shlapobersky, A.~Chen, and
  E.~Schneidman}, {\em High-order social interactions in groups of mice},
  Elife, 2 (2013), p.~e00759.

\bibitem{shlens2009structure}
{\sc J.~Shlens, G.~D. Field, J.~L. Gauthier, M.~Greschner, A.~Sher, A.~M.
  Litke, and E.~Chichilnisky}, {\em The structure of large-scale synchronized
  firing in primate retina}, The Journal of Neuroscience, 29 (2009),
  pp.~5022--5031.

\bibitem{shlens2006structure}
{\sc J.~Shlens, G.~D. Field, J.~L. Gauthier, M.~I. Grivich, D.~Petrusca,
  A.~Sher, A.~M. Litke, and E.~Chichilnisky}, {\em The structure of
  multi-neuron firing patterns in primate retina}, The Journal of neuroscience,
  26 (2006), pp.~8254--8266.

\bibitem{stricker2008fast}
{\sc J.~Stricker, S.~Cookson, M.~R. Bennett, W.~H. Mather, L.~S. Tsimring, and
  J.~Hasty}, {\em A fast, robust and tunable synthetic gene oscillator},
  Nature, 456 (2008), pp.~516--519.

\bibitem{tang2008maximum}
{\sc A.~Tang, D.~Jackson, J.~Hobbs, W.~Chen, J.~L. Smith, H.~Patel, A.~Prieto,
  D.~Petrusca, M.~I. Grivich, A.~Sher, et~al.}, {\em A maximum entropy model
  applied to spatial and temporal correlations from cortical networks in
  vitro}, The Journal of Neuroscience, 28 (2008), pp.~505--518.

\bibitem{vinje2000sparse}
{\sc W.~E. Vinje and J.~L. Gallant}, {\em Sparse coding and decorrelation in
  primary visual cortex during natural vision}, Science, 287 (2000),
  pp.~1273--1276.

\bibitem{wang2010review}
{\sc Z.~Wang, Y.~Ma, F.~Cheng, and L.~Yang}, {\em Review of pulse-coupled
  neural networks}, Image and Vision Computing, 28 (2010), pp.~5--13.

\bibitem{watanabe2013pairwise}
{\sc T.~Watanabe, S.~Hirose, H.~Wada, Y.~Imai, T.~Machida, I.~Shirouzu,
  S.~Konishi, Y.~Miyashita, and N.~Masuda}, {\em A pairwise maximum entropy
  model accurately describes resting-state human brain networks}, Nature
  communications, 4 (2013), p.~1370.

\bibitem{xu2016dynamical}
{\sc Z.-Q.~J. Xu, G.~Bi, D.~Zhou, and D.~Cai}, {\em A dynamical state
  underlying the second order maximum entropy principle in neuronal networks},
  Communications in Mathematical Sciences, 15 (2017), pp.~665--692.

\bibitem{zhou2013causal}
{\sc D.~Zhou, Y.~Xiao, Y.~Zhang, Z.~Xu, and D.~Cai}, {\em Causal and structural
  connectivity of pulse-coupled nonlinear networks}, Physical review letters,
  111 (2013), p.~054102.

\end{thebibliography}
 
\end{document}